\documentclass[twocolumn,showpacs,preprintnumbers,amsmath,amssymb,prb,superscriptaddress]{revtex4}
\usepackage{epsfig}
\usepackage{graphicx}
\usepackage{dcolumn}
\usepackage{bm}
\usepackage{color} 

\begin{document}

\title{
Optical study of charge instability in CeRu$_2$Al$_{10}$ in comparison with CeOs$_2$Al$_{10}$ and CeFe$_2$Al$_{10}$
}
\author{Shin-ichi Kimura}
\email{kimura@ims.ac.jp}
\affiliation{UVSOR Facility, Institute for Molecular Science, Okazaki 444-8585, Japan}
\affiliation{School of Physical Sciences, The Graduate University for Advanced Studies (SOKENDAI), Okazaki 444-8585, Japan}
\author{Takuya Iizuka}
\affiliation{School of Physical Sciences, The Graduate University for Advanced Studies (SOKENDAI), Okazaki 444-8585, Japan}
\author{Hidetoshi Miyazaki}
\altaffiliation[Present address: ]{Faculty of Engineering, Nagoya Institute of Technology, Nagoya 466-8555, Japan}
\affiliation{UVSOR Facility, Institute for Molecular Science, Okazaki 444-8585, Japan}
\author{Tetsuya Hajiri}
\affiliation{UVSOR Facility, Institute for Molecular Science, Okazaki 444-8585, Japan}
\author{Masaharu Matsunami}
\affiliation{UVSOR Facility, Institute for Molecular Science, Okazaki 444-8585, Japan}
\affiliation{School of Physical Sciences, The Graduate University for Advanced Studies (SOKENDAI), Okazaki 444-8585, Japan}
\author{Tatsuya Mori}
\altaffiliation{Present address: Graduate School of Pure and Applied Sciences, University of Tsukuba, Tsukuba 305-8577, Japan }
\affiliation{UVSOR Facility, Institute for Molecular Science, Okazaki 444-8585, Japan}
\author{Akinori Irizawa}
\affiliation{The Institute of Scientific and Industrial Research, Osaka University, Ibaraki, Osaka 567-0047, Japan}
\author{Yuji Muro}
\altaffiliation{Present address: Department of Liberal Arts and Sciences, Toyama Prefectural University, Toyama 939 - 0398, Japan}
\affiliation{Department of Quantum Matter, ADSM, Hiroshima University, Higashi-Hiroshima, Hiroshima 739-8530, Japan}
\author{Junpei Kajino}
\affiliation{Department of Quantum Matter, ADSM, Hiroshima University, Higashi-Hiroshima, Hiroshima 739-8530, Japan}
\author{Toshiro Takabatake}
\affiliation{Department of Quantum Matter, ADSM, Hiroshima University, Higashi-Hiroshima, Hiroshima 739-8530, Japan}
\affiliation{Institute for Advanced Materials Research, Hiroshima University, Higashi-Hiroshima, Hiroshima 739-8530, Japan}
\date{\today}
\begin{abstract}
The anisotropic electronic structure responsible for the antiferromagnetic transition in CeRu$_2$Al$_{10}$ at the unusually high temperature of $T_0$~=~28~K was studied using optical conductivity spectra, Ce~$3d$ X-ray photoemission spectra, and band calculation.
It was found that the electronic structure in the $ac$ plane is that of a Kondo semiconductor, whereas that along the $b$ axis has a nesting below 32~K (slightly higher than $T_0$).
These characteristics are the same as those of CeOs$_2$Al$_{10}$ [S. Kimura {\it et al.}, Phys. Rev. Lett. {\bf 106}, 056404 (2011).].
The $c$-$f$ hybridization intensities between the conduction and $4f$ electrons of CeRu$_2$Al$_{10}$ and CeOs$_2$Al$_{10}$ are weaker than that of CeFe$_2$Al$_{10}$, showing no magnetic ordering.
These results suggest that the electronic structure with one-dimensional weak $c$-$f$ hybridization along the $b$ axis combined with two-dimensional strong hybridization in the $ac$ plane causes charge-density wave (CDW) instability, and the CDW state then induces magnetic ordering.
\end{abstract}

%
%
%
%
\maketitle
%
\section{Introduction}
%
\begin{figure}[b]
\begin{center}
\includegraphics[width=0.40\textwidth]{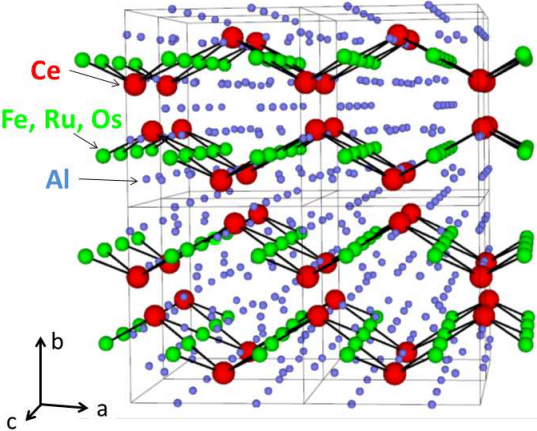}
\end{center}
\caption{
(Color online)
Crystal structure of Ce$M_2$Al$_{10}$ ($M=$Fe, Ru, Os).
Eight unit cells marked by lines are depicted.
}
\label{Crystal}
\end{figure}
Kondo semiconductors/insulators (KIs) such as SmB$_6$, YbB$_{12}$, Ce$_3$Bi$_4$Pt$_3$, and CeRhSb have a tiny energy gap at the Fermi level ($E_F$) due to strong hybridization between the conduction electrons and mostly localized $f$ electron; namely, $c$-$f$ hybridization.~\cite{Aeppli1992,Takabatake1998}
These KIs do not necessarily exhibit magnetic ordering throughout a given temperature region.
The recently discovered KIs Ce$M_2$Al$_{10}$ ($M$~=~Os, Ru), 
which crystallize in an orthorhombic YbFe$_2$Al$_{10}$-type crystal structure (space group $Cmcm$, No. 63CFig.~\ref{Crystal}),~\cite{Thiede1998} however, show an anomalous antiferromagnetic phase transition at rather high $T_0$ temperatures (28.7~K for $M$~=~Os, 27.3~K for Ru), which are lower than the Kondo temperatures ($T_{\rm K}\sim$100~K for Os, 60~K for Ru).~\cite{Strydom2009,Nishioka2009,Matsumura2009}
Recent studies using neutron diffraction and $\mu$SR have revealed that the magnetic structure has a propagation vector of $q=(0,1,0)$ and collinear antiferromagnetic moments orienting along the $c$ axis.~\cite{Khalyavin2010, Adroja2010, Robert2010, Kambe2010, Kato2011} 
Since the Ce--Ce distance is longer than 5~\AA and Gd-based counterparts of GdRu$_2$Al$_{10}$ and GdOs$_2$Al$_{10}$ have a lower N\'eel temperature $T_N$, at which the phase transition can be solely attributed to the Ruderman-Kittel-Kasuya-Yoshida (RKKY) interaction, the RKKY interaction is not believed to be the origin of the phase transition at $T_0$.~\cite{Nishioka2009, Muro2011}
The origin of the magnetic ordering, therefore, remains to be addressed.

So far, we have investigated the temperature dependence of the electronic structure of CeOs$_2$Al$_{10}$ along the three principal axes using the optical conductivity [$\sigma(\omega)$] spectra.~\cite{Kimura2011-1}
As a result, it has been found that energy gaps along the $a$ and $c$ axes open at temperatures higher than $T_0$ and that the gap sizes monotonically increase on cooling.
Because the $\sigma(\omega)$ spectra in the $ac$ plane are consistent with the $\sigma(\omega)$ spectra of conventional KIs reported previously, the electronic structure in the $ac$ plane can be regarded as that of a KI.
Along the $b$ axis, however, an energy gap is suddenly formed at $T_0$.
Since the shape of the energy gap along the $b$ axis is similar to that of charge-density wave (CDW)/spin-density wave (SDW) compounds reported previously, the origin has been attributed to band nesting due to charge instability.
The change in the $\sigma(\omega)$ spectrum begins at a slightly higher temperature than $T_0$, and the energy gap structure clearly appears below $T_0$.
Based on these observations, we proposed that a change in electronic structure due to CDW instability induces the antiferromagnetic ordering at $T_0$.

In another KI, CeFe$_2$Al$_{10}$, with no magnetic ordering, the Kondo effect appears anisotropically.~\cite{Muro2009,Muro2010-1}
The electronic structure along the $b$ axis is different from the isotropic electronic structure in the $ac$ plane.~\cite{Kimura2011-2}
The Kondo temperature $T_K$ can be evaluated using the temperature-dependent $\sigma(\omega)$ spectra and the temperature of maximum electrical resistivity $T_{max}$.
Thus, the evaluated intensity of the Kondo effect along the $b$ axis is weaker than that in the $ac$ plane.

In this paper, we focus on the relationship between the anisotropic Kondo effect and the anomalous magnetic ordering at $T_0$.
We show the anisotropic $\sigma(\omega)$ spectra of CeRu$_2$Al$_{10}$, to clarify the anisotropic electronic structure.
In CeOs$_2$Al$_{10}$, the peak structure at a photon energy $\hbar\omega$ of $\sim$20~meV appears below 38~K, which is the same temperature of maximum magnetic susceptibility $\chi(T)$.
The $\sigma(\omega)$ spectrum of CeRu$_2$Al$_{10}$ also has a shoulder structure at $\hbar\omega\sim$20~meV, as does CeOs$_2$Al$_{10}$.
The shoulder appears at 30~K, which agrees with the temperature of maximum $\chi(T)$.
The temperature dependence of the intensity obeys the square of the gap function $\Delta(T)^2$, which is consistent with the BCS theory as in the case of CeOs$_2$Al$_{10}$.
This fact suggests that the energy gap along the $b$ axis originates from the pairing of conduction electrons, namely charge instability or the CDW state, and that this in turn induces the magnetic ordering.
We point out that the creation of the CDW state is strongly related to the $c$-$f$ hybridization intensity.

\section{Experimental and calculation procedure}
Single-crystalline Ce$M_2$Al$_{10}$s ($M=$Fe, Ru, Os) were synthesized by the Al-flux method~\cite{Muro2010-2} and the surfaces were well-polished using 0.3~$\mu$m grain-size Al$_{2}$O$_{3}$ lapping film sheets for the optical reflectivity [$R(\omega)$] measurements.
Near-normal incident polarized $R(\omega)$ spectra were acquired in a very wide photon-energy region of 2~meV -- 30~eV to ensure accurate Kramers-Kronig analysis (KKA).
Martin-Puplett and Michelson type rapid-scan Fourier spectrometers (JASCO Co. Ltd., FARIS-1 and FTIR610) were used at the photon energy $\hbar\omega$ regions of 2~--~30~meV and 5~meV~--~1.5~eV, respectively, with a specially designed feed-back positioning system to maintain the overall uncertainty level less than $\pm$0.5~\% in the $T$ range of 10~--~300~K.~\cite{Kimura2008}
To obtain the absolute $R(\omega)$ values, the samples were evaporated {\it in-situ} with gold, whose spectrum was then measured as a reference.
At $T=300$~K, $R(\omega)$ was measured for energies 1.2--30~eV by using synchrotron radiation.~\cite{Fukui2001}
In order to obtain $\sigma(\omega)$ via KKA of $R(\omega)$, the spectra were extrapolated below 2~meV with a Hagen-Rubens function, and above 30~eV with a free-electron approximation $R(\omega) \propto \omega^{-4}$.~\cite{DG}

To evaluate the mean valence of Ce ion in Ce$M_2$Al$_{10}$, an X-ray photoemission spectroscopy (XPS) experiment was performed on the Ce $3d$ core levels using an Mg $K\alpha$ X-ray source ($h\nu$=1253.6~eV) and a 100~mm-radius hemispherical photoelectron analyzer (VG Scienta SES-100).
The base pressure of the chamber was less than $2\times10^{-8}$~Pa.
The sample temperature and total energy resolution were set at approximately 10~K and 0.6~eV, respectively.
Clean sample surfaces were prepared inside an ultra high-vacuum chamber by scraping with a diamond filler.
After cleaning, the oxygen and carbon contaminations were checked by monitoring the intensity of the O~$1s$ and C~$1s$ photoemission peaks maintained within the noise level during the measurement.

Local-density approximation (LDA) band structure calculation was performed by the full potential linearized augmented plane wave plus local orbital (LAPW+lo) method including spin-orbit interaction (SOI) implemented in the {\sc Wien2k} code.~\cite{WIEN2k} 
The lattice parameters reported in the literature~\cite{Thiede1998} were used for the calculation.
$R_{MT}K_{max}$ (the smallest MT radius multiplied by the maximum $k$ value in the expansion of plane waves in the basis set), which determines the accuracy of the basis set used, was set at 7.0.
The total $k$ number in a Brillouin zone was sampled with 10~000~$k$-points.
Note that the LDA band calculation produces a metallic band structure, which is not consistent with the experimental KI character.
However, since the purpose of this study was to investigate the $c$-$f$ hybridization intensity as well as the mean valence number from first-principle band calculation, we used the metallic band structure derived from the LDA calculation.

\section{Results and Discussion}
\subsection{Anisotropic $\sigma(\omega)$ of CeRu$_2$Al$_{10}$}
%
\begin{figure}[b]
\begin{center}
\includegraphics[width=0.40\textwidth]{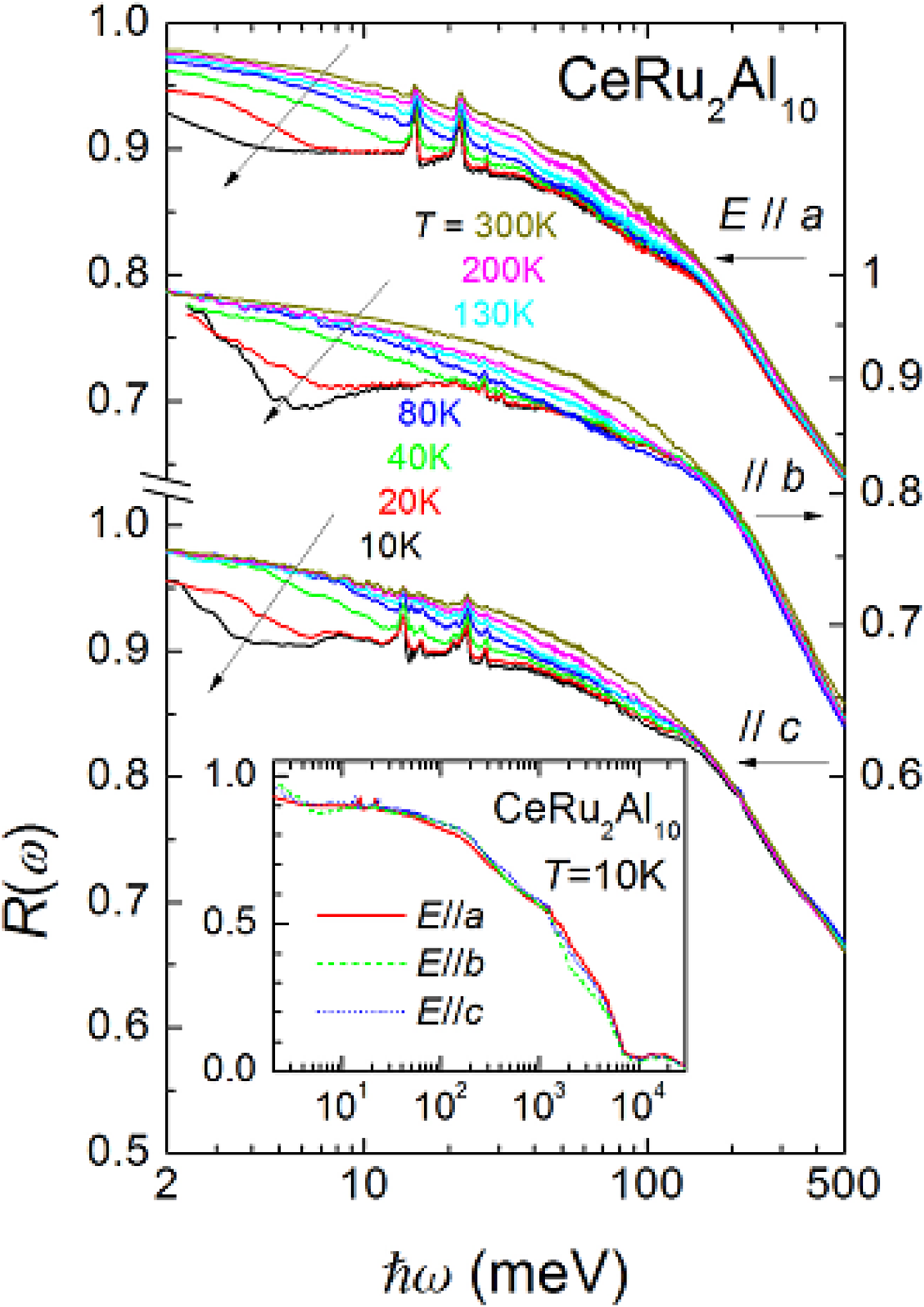}
\end{center}
\caption{
(Color online)
Low-energy portion of the temperature-dependent polarized reflectivity [$R(\omega)$] spectra of CeRu$_2$Al$_{10}$ along the three principal axes.
(Inset) The entire reflectivity spectra at 10~K along the $a$ axis (solid lines), $b$ axis (dashed lines), and $c$ axis (dotted lines).
}
\label{reflectivity}
\end{figure}
The obtained $R(\omega)$ spectra of CeRu$_2$Al$_{10}$ along the $a$ axis ($E\|a$), $b$ axis ($E\|b$), and $c$ axis ($E\|c$) are shown in Fig.~\ref{reflectivity}.
As can be seen in the inset, the $R(\omega)$ spectra monotonically decrease up to $\hbar\omega\sim$~10~eV, because the conduction band of aluminium expands to about 10~eV below $E_{\rm F}$ as shown in the band calculation in Fig.~\ref{DOS}(b).
In the middle-infrared region from 100 to 500~meV, there are small shoulder structures.
This shape is not a characteristic double-peak structure with energy splitting of 250~meV, which originates from optical excitation from the occupied state at $E_F$ to the unoccupied Ce~$4f$ state with spin-orbit splitting usually seen in itinerant Ce compounds.~\cite{Kimura2009}
This indicates that the hybridization of the Ce~$4f$ state with other states is not so large.

Let us focus on the spectra below 100~meV.
In the range of 10--30~meV at 10~K, there are sharp peaks due to optical phonons: three sharp peaks in $E\|a$, three small peaks in $E\|b$, and five peaks in $E\|c$.
Except for these peaks, the $R(\omega)$ spectra for all principal axes at 300~K are Drude-like spectra that increase to unity with decreasing photon energy, indicating a metallic character.
Below 80~K, the value of $R(\omega)$ below 150~meV rapidly decreases on cooling, and eventually, other metallic $R(\omega)$ spectra appear along all axes near the lowest accessible photon energy of 2~meV.
This indicates that the normal-metallic state changes to the low-carrier metallic state on cooling.
This result is consistent with the temperature dependence of electrical resistivity.~\cite{Nishioka2009}

\begin{figure}[b]
\begin{center}
\includegraphics[width=0.40\textwidth]{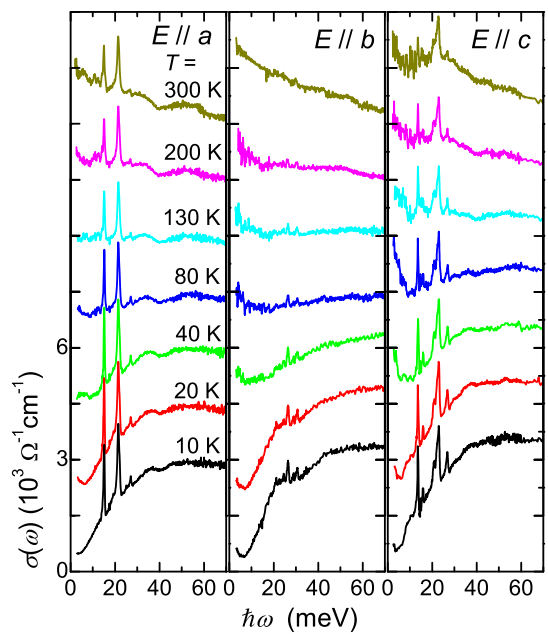}
\end{center}
\caption{
(Color online)
Temperature-dependent polarized optical conductivity [$\sigma(\omega)$] spectra of CeRu$_2$Al$_{10}$ in the photon energy $\hbar\omega$ region below 70~meV.
Each of the lines is shifted by 1.5$\times10^3$~$\Omega^{-1}$cm$^{-1}$ for clarity.
}
\label{OC}
\end{figure}
The temperature-dependent $\sigma(\omega)$ spectra derived from KKA of the $R(\omega)$ spectra in Fig.~\ref{reflectivity} are shown in Fig.~\ref{OC}.
The $\sigma(\omega)$ spectra for all principal axes at 300~K monotonically increase with decreasing photon energy, again indicating a metallic character.
A common feature of all of the principal axes is that the $\sigma(\omega)$ intensity below $\hbar\omega$~=~60~meV decreases on cooling, and at 80~K, a broad shoulder structure appears at about 35~meV, which suggests the existence of an energy gap due to strong hybridization in a similar manner to other KIs.~\cite{Kimura1994,Bucher1994,Okamura1998,Matsunami2003}

In $E\|a$ and $E\|c$, gentle shoulder structures at about 35~meV gradually evolve below 80~K and energy gaps appear at the lower energy side of the peak as the temperature decreases to 10~K.
Unlike in $E\|b$, however, neither the gap shape nor the energy change with temperature (except for the reduction in thermal broadening).
This implies that the intensity below the energy gap does not transfer only to the region immediately above the energy gap, but to a wide energy range.
This temperature-dependent gap structure is consistent with that of other KIs reported previously;~\cite{Kimura1994,Bucher1994,Okamura1998,Matsunami2003} namely, the temperature-dependent $c$-$f$ hybridization gap.
Therefore, the electronic structures in $E\|a$ and $E\|c$ are similar to those of conventional KIs.

On the other hand, in $E\|b$, another shoulder structure evolves at 20~meV below 20~K, which is similar to that of CeOs$_2$Al$_{10}$.~\cite{Kimura2011-1}
This suggests the same origin of the shoulder structure in the two systems.
The shoulder structure emerges below $T\sim$~80~K, developing at 35~meV due to the $c$-$f$ hybridization gap.
This characteristic is also consistent with that of CeOs$_2$Al$_{10}$.

Below 40~K, the tails of a Drude structure appear at around 5~meV in all axes with the shape differing from the spectral shape above 80~K, which is close to the $T_K$ of 60~K.
According to the electrical resistivity data, the direct-current conductivity is about 1000~$\Omega^{-1}$cm$^{-1}$ or higher in all of the axes.
As a result, very narrow Drude peaks suggesting quasi-particles with a low scattering rate can be expected to appear below 5~meV.
This implies that the character of the carriers changes from a normal metal to quasi-particles on cooling through $T_K$, as in the case of CeFe$_2$Al$_{10}$ and other heavy fermion materials.~\cite{Kimura2011-2, Iizuka2010}

\subsection{$c$-$f$ hybridization intensity in $E\|a$ and $E\|c$}
\begin{figure}[b]
\begin{center}
\includegraphics[width=0.35\textwidth]{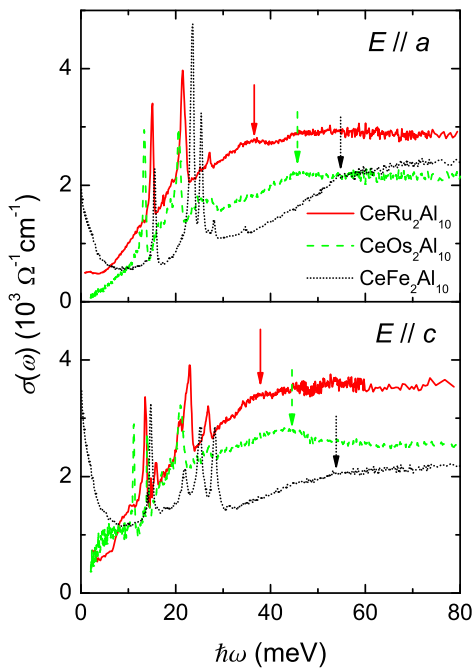}
\end{center}
\caption{
(Color online)
Optical conductivity [$\sigma(\omega)$] spectra of Ce$M_2$Al$_{10}$ ($M=$Fe, Ru, Os) along the $a$ axis (upper) and $c$ axis (lower) in the $\hbar\omega$ range below 80~meV at $T=$10~K.
The downward arrows indicate the absorption structures in the $c$-$f$ hybridization gaps.
}
\label{EnergyGap}
\end{figure}
Let us discuss the $c$-$f$ hybridization gap in $E\|a$ and $E\|c$ for the three systems of Ce$M_2$Al$_{10}$ ($M$~=~Ru, Os, Fe).
In Fig.~\ref{EnergyGap}, shoulder structures exist as indicated by the arrows at 35, 45, and 55 meV for $M$~=~Ru, Os, and Fe, respectively.
In CeFe$_2$Al$_{10}$, the shoulder structure is ascribed to the optical transition between the bonding and antibonding states of the $c$-$f$ hybridization band.~\cite{Kimura2011-2}
According to the periodic Anderson model that describes the $c$-$f$ hybridization band, the minimum energy difference between the bonding and antibonding bands at the same wave vector $k$ point is roughly double the effective hybridization energy $\tilde{V}$.~\cite{Hewson1993}
Because the energy of the shoulder structure is related to $\tilde{V}$, the effective $c$-$f$ hybridization energy increases as $M$ changes from Ru to Os and Fe.

\begin{figure}[b]
\begin{center}
\includegraphics[width=0.40\textwidth]{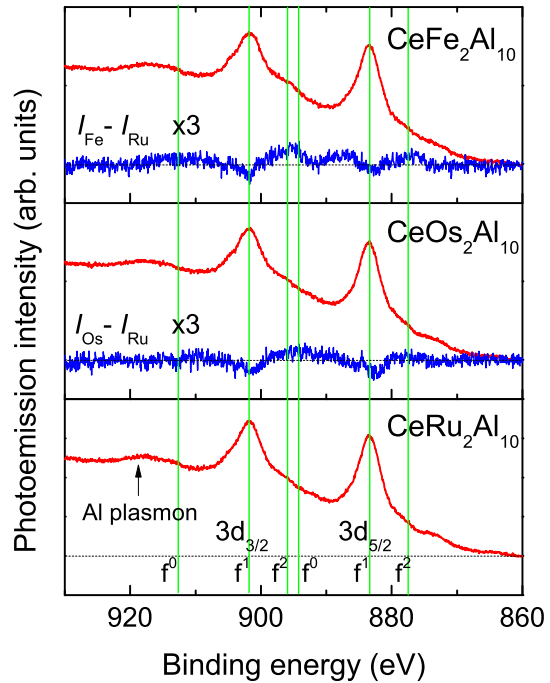}
\end{center}
\caption{
(Color online)
Ce~$3d$ XPS spectra of Ce$M_2$Al$_{10}$ ($M=$Fe, Os, Ru) at 10~K.
The energies of the $f^0$, $f^1$, and $f^2$ final states of the $3d_{3/2}$ and $3d_{5/2}$ core levels are marked by vertical lines.
The subtraction spectra obtained by subtracting CeRu$_2$Al$_{10}$ from CeFe$_2$Al$_{10}$ and CeOs$_2$Al$_{10}$ are also plotted as $I_{\rm Fe}-I_{\rm Ru}$ and $I_{\rm Os}-I_{\rm Ru}$, respectively.
}
\label{Ce3dXPS}
\end{figure}
The mean valence of Ce ion can be evaluated from the Ce~$3d$ core level photoemission spectra (Ce~$3d$~XPS).~\cite{GS1983}
Figure~\ref{Ce3dXPS} shows the normalized Ce~$3d$~XPS spectra of Ce$M_2$Al$_{10}$ at 10~K.
However, conventional fitting of the $f^0$, $f^1$, and $f^2$ final states is difficult in this case, because a large plasmon peak of aluminium exists at about 16~eV higher on the binding energy side of the main peaks of $3d_{5/2}$ and $3d_{3/2}$.
The hybridization intensities of CeFe$_2$Al$_{10}$ and CeOs$_2$Al$_{10}$ were therefore estimated by subtraction; i.e., [$I_{\rm Fe}-I_{\rm Ru}$] and [$I_{\rm Os}-I_{\rm Ru}$], respectively.
These differential spectra show that the $f^1$ peak intensity decreases and $f^0$ peak intensity increases in the order of Ru$\rightarrow$Os$\rightarrow$Fe.
The mean valence $n_f$ can be roughly obtained by the function $n_f=1-I(f^0)/[I(f^0)+I(f^1)+2\times I(f^2)]$, where $I(f^n)$ is the integrated intensity of an $f^n$ peak.~\cite{Klein2009}
The value of $n_f$ decreases from 1.0 as $M$ changes from Ru to Os and Fe.
This suggests that the $c$-$f$ hybridization intensity increases in the order of Ru$\rightarrow$Os$\rightarrow$Fe.
This trend is consistent with the order of the energy gap size in these materials observed by $\sigma(\omega)$ spectra.
Note that the spectral feature of CeOs$_2$Al$_{10}$ did not change with temperature across $T_0$.
This implies that the mean valence does not change at $T_0$.

\begin{figure}[b]
\begin{center}
\includegraphics[width=0.40\textwidth]{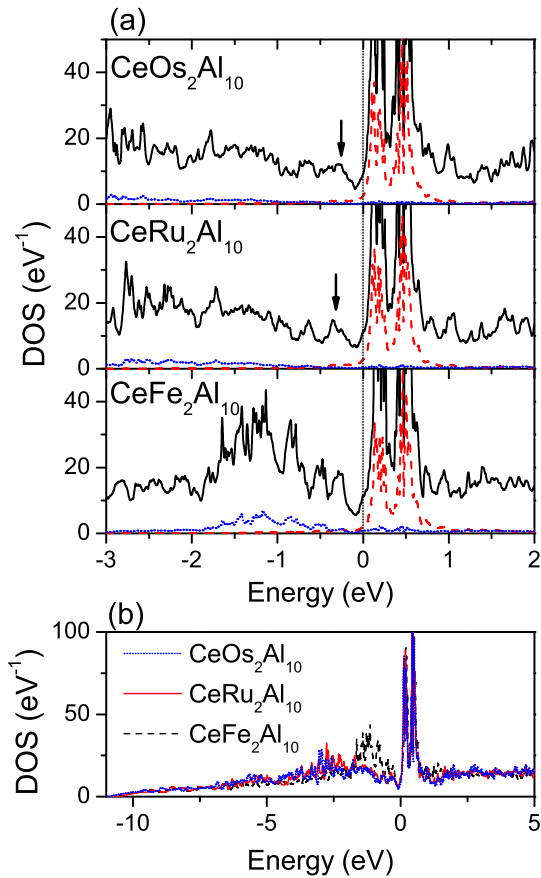}
\end{center}
\caption{
(Color online)
(a) Total density of states (DOS) of Ce$M_2$Al$_{10}$ ($M=$~Os, Ru, Fe) and partial DOS of the Ce~$f$ (dashed lines) and $M d$ (dotted lines) states near the Fermi level at an energy of 0~eV.
The shallowest peaks in the valence bands of CeRu$_2$Al$_{10}$ and CeOs$_2$Al$_{10}$ are denoted by downward arrows.
(b) Total DOS of Ce$M_2$Al$_{10}$ in the wide energy range.
}
\label{DOS}
\end{figure}
The Ce~$3d$ XPS result was confirmed by LDA band calculation.
The total density of states (DOS) and partial DOS of the Ce~$f$ and $M d$ states derived from the band calculation with SOI are shown in Fig.~\ref{DOS}(a).
The total DOS of all Ce$M_2$Al$_{10}$s in the wide energy range, shown in Fig.~\ref{DOS}(b), indicates that the structure near $E_F$ is overlaid on a large Al~$3s$-$3p$ conduction band with a bottom of about $-10$~eV.
The double sharp peak structure immediately above $E_F$ indicates the Ce~$4f_{7/2}$ and $4f_{5/2}$ states with SOI.
The $M d$ states of all materials that expand below $E_F$ differ from one another.
The Fe~$3d$ state of CeFe$_2$Al$_{10}$ is localized from $E_F$ to $-2$~eV, whereas the Ru~$4d$ state of CeRu$_2$Al$_{10}$ and the Os~$5d$ state widely expand in the valence band.
This suggests that hybridization of the Fe~$3d$ state to the Ce~$4f$ state is stronger than that in the case of Ru~$4d$ and Os~$5d$.
The shallowest peak in the occupied state of CeOs$_2$Al$_{10}$ is slightly closer than that of CeRu$_2$Al$_{10}$, as indicated by the arrows in the figure.
The peak originates from the $M d$ state; i.e., the Os~$5d$ state is located closer to $E_F$ than the Ru~$4d$ state due to the strong SOI.
A similar trend has been observed in alkaline-earth-filled skutterudites $AM_4$Sb$_{12}$ ($A$~=~alkaline-earth, $M$~=~Fe, Ru, Os).~\cite{Kimura2007}

To compare the $c$-$f$ hybridization intensity among the three compounds, the mean $4f$ electron number $n_f$ was evaluated from the band calculation.
As a result, $n_f$ was found to be 0.943 in CeRu$_2$Al$_{10}$ (most localized), 0.942 in CeOs$_2$Al$_{10}$, and 0.937 in CeFe$_2$Al$_{10}$ (most itinerant).
This variation is consistent with the order of the energy gap size in the $\sigma(\omega)$ spectra and the results obtained for the Ce~$3d$ XPS spectra, in which the $c$-$f$ hybridization intensity increases in the order of Ru$\rightarrow$Os$\rightarrow$Fe.
Therefore, the difference in the $c$-$f$ hybridization intensity originates from the different $M d$ states.

Note that the band calculation showed no energy gap structure, which is inconsistent with the observed $\sigma(\omega)$ spectra.
In conventional KIs, a semiconducting or semi-metallic band structure appears in the band calculation, although the energy scale is renormalized by a strong many-body interaction.
However, neither a gap nor a semi-metallic structure appears even in CeFe$_2$Al$_{10}$, which can be regarded as a conventional KI.
The reason for this inconsistency is not yet clear, but the evolution of the energy gap might be assisted by another mechanism.

All of the results of the edge in the $\sigma(\omega)$ spectra, the Ce~$3d$~XPS, and the band calculation indicate that the $c$-$f$ hybridization intensity increases in the order of CeRu$_2$Al$_{10}$ $\rightarrow$ CeOs$_2$Al$_{10}$ $\rightarrow$ CeFe$_2$Al$_{10}$.
This result is consistent with the reported pressure-dependent electrical resistivity $\rho(T)$.~\cite{Nishioka2009}
Applying pressure to cerium compounds is equivalent to increasing the $c$-$f$ hybridization intensity.
When pressure is applied to CeRu$_2$Al$_{10}$, $\rho(T)$ becomes similar to that of CeOs$_2$Al$_{10}$ at about 1.5~GPa, and above 4~GPa, $T_0$ disappears and $\rho(T)$ becomes similar to that of CeFe$_2$Al$_{10}$.
In the case of CeOs$_2$Al$_{10}$, $T_0$ also disappears above 2~GPa and $\rho(T)$ changes to that of CeFe$_2$Al$_{10}$.~\cite{Umeo2011}

At ambient temperature and pressure, the volumes of the unit cell are 0.836~nm$^3$ for CeFe$_2$Al$_{10}$, 0.857~nm$^3$ for CeRu$_2$Al$_{10}$, and 0.858~nm$^3$ for CeOs$_2$Al$_{10}$.~\cite{Nishioka2009}
This order is the same as the order of ionic radii of Fe, Ru, and Os.
The order of the $c$-$f$ hybridization intensities, however, is revised between CeRu$_2$Al$_{10}$ and CeOs$_2$Al$_{10}$.
This is considered to originate from the fact that the SOI of $5d$ electron is larger than that of $4d$.
The Os~$5d_{5/2}$ level is therefore located close to $E_F$, and is readily mixed to the Ce~$4f$ state.
As a result, the $c$-$f$ hybridization intensity of CeOs$_2$Al$_{10}$ becomes stronger than that of CeRu$_2$Al$_{10}$.

\subsection{Charge-density wave in $E\|b$}
%
\begin{figure}[b]
\begin{center}
\includegraphics[width=0.35\textwidth]{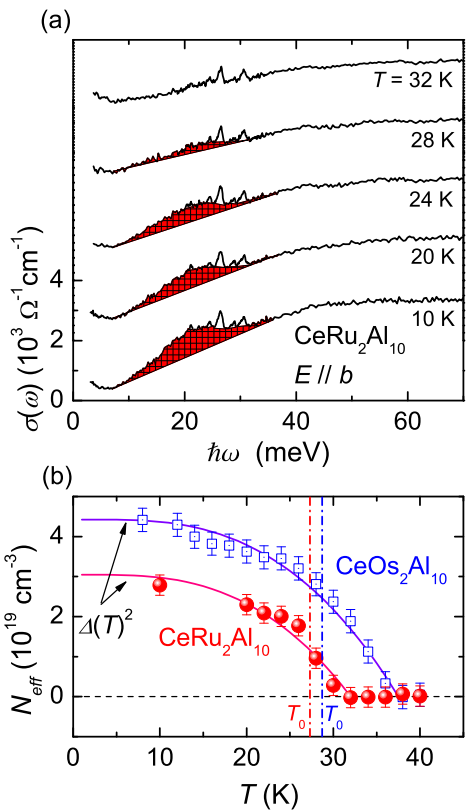}
\end{center}
\caption{
(Color online)
(a) Temperature-dependent optical conductivity [$\sigma(\omega)$] spectrum of CeRu$_2$Al$_{10}$ in $E\|b$.
The cross-hatched areas are shoulder structures caused by band nesting due to charge-density wave formation.
(b) Effective electron number ($N_{eff}$) of the shoulder structures shown in (a) plotted as a function of temperature.
The $N_{eff}$ of CeOs$_2$Al$_{10}$ is also plotted.
The solid lines indicate the square of the order parameter $\Delta(T)$ of the BCS theory.
The critical temperature $T^*$ of $\Delta(T)$ was set at 32~K for CeRu$_2$Al$_{10}$ and 37.5~K for CeOs$_2$Al$_{10}$.
The vertical dotted-dashed lines indicate the antiferromagnetic ordering temperature $T_0$ of CeRu$_2$Al$_{10}$ and CeOs$_2$Al$_{10}$.
}
\label{GapFunction}
\end{figure}
As shown in Fig.~\ref{OC}, the $\sigma(\omega)$ spectra in $E\|b$ have a shoulder structure at $\hbar\omega=$20~meV (referred to as the ``20~meV peak'' hereafter) at low temperatures.
To trace the temperature dependence in detail, the $\sigma(\omega)$ spectra were measured below 40~K in steps of 2~K.
Figure~\ref{GapFunction}(a) shows selected $\sigma(\omega)$ spectra in $E\|b$ at temperatures from 32 to 10~K.
On cooling below 28 K, the 20~meV peak extends above the background originating from the $c$-$f$ hybridization gap and optical phonon peaks appear at around 30~meV.
Because the shape of the $c$-$f$ hybridization gap in $E\|a$ and $E\|c$ is almost linear with the energy, the background in $E\|b$ was assumed to be linear between 8 and 35 meV.
The 20~meV peak intensity was then extracted by subtracting the background and the phonon peaks.
The obtained 20~meV peak intensities are shown as the cross-hatched areas in Fig.~\ref{GapFunction}(a).
To evaluate the total intensity, the effective electron number ($N_{eff}$) was calculated by using the function, 
\[
N_{eff} = \frac{4 m_0}{h^2 e^2}\int^{\infty}_{0}\sigma(\hbar\omega)d\hbar\omega,
\]
where $h$ is the Planck constant, $e$ the elementary charge, and $m_{0}$ the electron rest mass.~\cite{DG}

The obtained $N_{eff}$s for CeRu$_2$Al$_{10}$ and CeOs$_2$Al$_{10}$ are plotted as a function of temperature in Fig.~\ref{GapFunction}(b).
The solid curves show the square of the order-parameter $\Delta(T)$ given by the BCS theory, which is phenomenologically approximated as $\Delta(T)=\Delta_0\times \tanh{[1.74\times\surd(T_c/T-1)]}$.~\cite{Gruner1988}
All data points fall on the BCS function.
Consequently, it is strongly suggested that the charge carriers undergo pairing; i.e., charge instability or a CDW/SDW is formed along the $b$ axis.
This finding is essentially the same that obtained in previous studies of CDW and SDW transitions in $1T$-TiSe$_2$~\cite{Li2007}, (TMTSF)$_2$PF$_6$,~\cite{Dressel1997} and others.
The superlattice reflections of CeOs$_2$Al$_{10}$ observed along the $[0\bar{1}1]$ direction by electron diffraction are also consistent with the CDW/SDW scenario.~\cite{Muro2010-2}

By the fitting of $\Delta(T)^2$ with the BCS function, the critical temperature $T^*$ of CeRu$_2$Al$_{10}$ was determined to be 32~K and that of CeOs$_2$Al$_{10}$ to be 37.5~K.
The value of $T_c$ is certainly higher than that of $T_0$, but close to the temperature of maximum $\chi(T)$ ($\sim$30~K for CeRu$_2$Al$_{10}$, $\sim$40~K for CeOs$_2$Al$_{10}$).~\cite{Nishioka2009,Muro2010-2}
This implies that the 20~meV peak is related to the mechanism that determines the maximum of $\chi(T)$.
For example, in quasi one-dimensional conductors such as (TaSe$_4$)$_2$I, K$_{0.3}$MoO$_3$, TaSe$_4$I$_{0.5}$, and TaS$_3$, CDW fluctuation occurs when the mean-field transition temperature $T_c^{MF}$ exceeds the CDW transition temperature $T_c$.~\cite{Johnston1985}
$\chi(T)$ as well as the electronic structure near $E_F$ can be explained by CDW fluctuation based on the mean-field theory.~\cite{Lee1973}
At $T_c^{MF}$, the magnetic spin susceptibility begins to decrease from a constant Pauli spin susceptibility at higher temperatures, and the DOS at $E_F$ begins to decrease while the intensity shifts to the gap edge due to band nesting.
If $T^*$ and $T_0$ of CeRu$_2$Al$_{10}$ and CeOs$_2$Al$_{10}$ are regarded as $T_c^{MF}$ and $T_c$, respectively, the spectral change and the temperature dependence of the magnetic susceptibility of there materials can be explained by CDW fluctuation based on the mean-field theory.
This result strongly suggests that the energy gaps along the $b$ axis of CeRu$_2$Al$_{10}$ and CeOs$_2$Al$_{10}$ originate from CDW formation.

In CeFe$_2$Al$_{10}$, $\chi(T)$ also reaches its maximum at around 70~K, although there is no magnetic transition.
The $\sigma(\omega)$ spectra have no peak structure below the $c$-$f$ hybridization gap energy of 55~meV.
Therefore, the maximum of $\chi(T)$ is considered to originate from the exponential decrease in the carrier density, as found in other KIs.~\cite{Aeppli1992,Takabatake1998}

To summarize the results of the present work and our previous studies,~\cite{Kimura2011-1,Kimura2011-2} CeOs$_2$Al$_{10}$ and CeRu$_2$Al$_{10}$ with a CDW electronic structure along the $b$ axis show magnetic ordering at $T_0$, whereas CeFe$_2$Al$_{10}$ without such an electronic structure does not have magnetic ordering.
This relation strongly suggests that the formation of the CDW along the $b$ axis triggers the magnetic ordering.
This conjecture is consistent with the pressure-dependent electrical resistivity data of CeOs$_2$Al$_{10}$.~\cite{Umeo2011}
Under pressure, $T_0$ suddenly disappears at the critical pressure ($\sim$2.5~GPa) at which the evaluated activation energy ($\Delta_L$) becomes 0 and the energy gap is closed.
The energy gap $2\Delta_L$ at ambient pressure is about 3~meV, which is similar to the onset energy ($\sim$5~meV) of the energy gap observed in the $\sigma(\omega,T)$ spectrum at 10~K devided by that at 40~K.~\cite{Kimura2011-1}
The magnetic ordering disappears above $T_0$, at which the onset energy becomes 0 and the energy gap is closed.
Therefore, both the results of the $\sigma(\omega)$ spectrum and the pressure-dependent electrical resistivity indicate that the energy gap formation above $T_0$ induces the magnetic ordering.

As discussed above, the $c$-$f$ hybridization intensity increases in the order of CeRu$_2$Al$_{10}$ $\rightarrow$ CeOs$_2$Al$_{10}$ $\rightarrow$ CeFe$_2$Al$_{10}$.
The character of all three systems in the $ac$ plane is that of a KI because of the stronger $c$-$f$ hybridization intensity in the $ac$ plane than along the $b$ axis.
Along the $b$ axis, CeFe$_2$Al$_{10}$, which has the strongest hybridization intensity, behaves as a KI, but the electronic states of CeRu$_2$Al$_{10}$ and CeOs$_2$Al$_{10}$, which have weak hybridization intensity, become the CDW state.
This suggests that the CDW transition originates from one-dimensional weak $c$-$f$ hybridization combined with two-dimensional strong hybridization.
In this sense, CeRu$_2$Al$_{10}$ and CeOs$_2$Al$_{10}$ are in a new class of KIs in which such strong anisotropy causes magnetic ordering at a high transition temperature.

\section{Conclusion}
The electronic structure of CeRu$_2$Al$_{10}$ was investigated using temperature-dependent polarized optical conductivity [$\sigma(\omega)$] spectra combined with Ce~$3d$ X-ray photoemission spectra (XPS) and band calculation.
Along the $a$ and $c$ axes, the spectral weight below 35~meV monotonically decreased on cooling, indicating a Kondo semiconducting character.
The temperature variation of the energy gap compared with those of CeFe$_2$Al$_{10}$ and CeOs$_2$Al$_{10}$, and their Ce~$3d$~XPS results, suggest that the intensity of $c$-$f$ hybridization between the conduction and $4f$ electrons increases in the order of CeRu$_2$Al$_{10}$, CeOs$_2$Al$_{10}$, and CeFe$_2$Al$_{10}$.
In contrast, a CDW energy gap opened below 32~K and 37.5~K along the $b$ axis in CeRu$_2$Al$_{10}$ and CeOs$_2$Al$_{10}$, respectively.
We conclude that the antiferromagnetic ordering at $T_0$ in CeRu$_2$Al$_{10}$ and CeOs$_2$Al$_{10}$ is assisted by CDW formation along the $b$ axis.

\section*{Acknowledgments}
We would like to thank UVSOR staff members for their technical support.
Part of this work was supported by the Use-of-UVSOR Facility Program (BL7B, 2009) of the Institute for Molecular Science.
The work was partly supported by a Grant-in-Aid for Scientific Research from MEXT of Japan (Grant No.~22340107, 20102004).

%
\end{document}